\documentclass[prb,twocolumn,showpacs]{revtex4}
\usepackage{amsmath}
\usepackage{amssymb}
\usepackage{graphicx}

\begin{document}

\title{Berry's phase and the anomalous velocity of Bloch wavepackets}

\author{Y.~D.~Chong}
\email{yidong.chong@yale.edu}

\affiliation{Department of Applied Physics, Yale University, New
  Haven, Connecticut 06520}

\date{\today}


\begin{abstract}
  The semiclassical equations of motion for a Bloch electron include an
  anomalous velocity term analogous to a $k$-space ``Lorentz force'', with the
  Berry connection playing the role of a vector potential.  By examining the
  adiabatic evolution of Bloch states in a monotonically-increasing vector
  potential, I show that the anomalous velocity can be explained as the
  difference in the Berry's phase acquired by adjacent Bloch states within a
  wavepacket.
\end{abstract}

\maketitle

When inversion or time-reversal symmetry is broken, the semiclassical
motion for a Bloch electron is known to contain an additional
non-vanishing term analogous to the Lorentz force in momentum-space:
\begin{equation}
  \dot{r} = \frac{1}{\hbar} \nabla_k E_k + \dot{k} \times \left(\nabla_k
    \times \mathcal{A}_k\right).
  \label{rdot}
\end{equation}
Here, $k$ is the reduced wave-vector, $E_k$ is the band energy, $\nabla_k$
denotes a $k$-space derivative, and $\mathcal{A}_k$ is defined by
\begin{equation}
  \vec{\mathcal{A}}_k = \frac{1}{i} \int_\Omega \! d^d\!r\, u_{k}^*(r)
  \nabla_{k} u_{k}(r),
  \label{Ak}
\end{equation}
with the integral taken over the unit cell $\Omega$ and $u_k(r)$ denoting the
Bloch function.  The ``anomalous velocity''---the second term on the right
hand side of (\ref{rdot})---was originally derived by Karplus and
Luttinger\cite{KL} in their explanation of the extraordinary Hall coefficients
of ferromagnetic materials, based on a careful examination of wavepacket
dynamics.  Subsequently, Chang and Niu re-derived the anomalous velocity using
an effective-Lagrangian technique, and pointed out that $\vec{\mathcal{A}}_k$
is the important quantity known as the Berry connection \cite{ChangNiu}.

In its original context, the Berry connection describes the gauge structure of
a quantum state as it undergoes adiabatic evolution, following a trajectory
$\vec{\lambda}(t)$ in some parameter space of the system Hamiltonian
$H(\lambda(t))$.  The line integral of the Berry curvature, taken over
$\vec{\lambda}(t)$, yields ``Berry's phase''---an additional phase acquired by
the quantum state during the adiabatic process \cite{Berry}.  As first
appreciated by Simon \cite{Simon}, the Berry connection also emerges within
Bloch systems, in a manner that appears to be quite different: the reduced
wave-vector $k$ serves as the ``parameter'' for the reduced Hamiltonian
$H(k)$, and $\vec{\mathcal{A}}_k$ describes the gauge structure of the Bloch
functions $u_k(r)$ within the Brillouin zone.  (In particular, the integral of
$\vec{\mathcal{A}}_k$ along the boundary of a two-dimensional Brillouin zone
yields the TKNN number, which equals the index of the integer quantum Hall
effect \cite{TKNN}.)

I would like to present a derivation of the anomalous velocity that clarifies
its relationship with adiabatic quantum evolution, the context in which the
Berry connection first arose.  One virtue of this derivation is that it
provides a simple geometrical explanation of why the anomalous velocity
involves the $k$-space curl of the Berry connection (the ``Berry curvature'').
The idea is simple: a small DC electric field can be represented by a constant
vector potential that increases monotonically with time.  Each Bloch state
undergoes adiabatic evolution in this vector potential, and acquires a Berry's
phase.  Each Bloch component of a wavepacket undergoes a different $k$-space
trajectory, and acquires a different Berry's phase.  The resulting Berry's
phase differences, characterized by the Berry curvature, conspire to induce
the anomalous term in the velocity of the wavepacket as a whole.

For simplicity of presentation, I will ignore the magnetic field
(whose presence does not alter the final results).  A Hamiltonian for
a Bloch electron subjected to an additional small electric field
$\mathcal{E} \hat{x}$ is
\begin{equation}
  H' = -\frac{\hbar^2}{2m}\nabla^2 + V(r) - e \mathcal{E} x,
\end{equation}
where $V(r)$ is the lattice potential, which can break inversion symmetry.  A
gauge transformation $\phi \rightarrow \phi - (1/c)\partial_t \Lambda$ and $A
\rightarrow A + \nabla \Lambda$, where $\Lambda = - \mathcal{E}cxt$, yields
the equivalent periodic time-dependent Hamiltonian
\begin{equation}
  H(t) = \frac{\hbar^2}{2m} \left[-i\nabla + \frac{e t}{\hbar}
    \vec{\mathcal{E}}\, \right]^2 + V(r).
  \label{H}
\end{equation}
The quantum states of the new Hamiltonian have an extra phase factor
of $\exp(-ie\mathcal{E}xt/\hbar)$, which is the same for all states
and therefore irrelevant.

As pointed out by Kittel\cite{Kittel}, $H$ now has the form of a reduced
Hamiltonian:
\begin{eqnarray}
  H(t) &=& H(\vec{q}(t)), \qquad \vec{q}(t) = \frac{et}{\hbar} \,
  \vec{\mathcal{E}}, \\
  H(k) &\equiv& \frac{\hbar^2}{2m} \left[-i\nabla + \vec{k} \right]^2 + V(r).
  \label{Hk}
\end{eqnarray}
This allows us to describe the effects of the electric field in terms of
adiabatic evolution.  Similar considerations were used by Zak in a related
one-dimensional model\cite{Zak}.

The Bloch states are eigenstates of $H(\vec{k})$ with band energies $E(k)$
(suppressing the irrelevant band index):
\begin{equation}
  H(k)\, u_k(r) = E(k)\, u_k(r).
\end{equation}
It is easily seen that
\begin{equation}
  H(k') \left[u_{k'+k}(r)\,e^{ikr}\right] = E(k'+k)\,
  \left[u_{k'+k}(r)\,e^{ikr}\right].
\end{equation}
Therefore, at each time $t$ the Hamiltonian $H(q(t))$ possesses a set
of instantaneous eigenstates $u_{q(t)+k}(r)\,e^{ikr}$.  (For $t=0$,
these are the usual Bloch states.)  If the electric field
$\mathcal{E}$ is weak, the change is adiabatic.

Suppose we have the initial state
\begin{equation}
  \psi(r,t=0) = u_k(r)\,e^{ikr}.
\end{equation}
According to Berry's theorem \cite{Berry}, the state at time $t$ is
\begin{equation}
  \psi(t) = u_{k(t)}\,e^{ikr} \, \exp\left[-\frac{i}{\hbar}\int_0^tdt'
    E(k(t'))\right]\, e^{i\gamma_k(t)},
  \label{psit}
\end{equation}
where $k(t) \equiv k + q(t)$.  Berry's phase, $\gamma_k(t)$, is given
by
\begin{eqnarray}
  \gamma_k(t) &=& i \int d\vec{q}' \cdot \left[\int_\Omega d^dr\,
    \left(u_{q'+k}e^{ikr}\right)^* \nabla_{q'}
    \left(u_{q'+k}e^{ikr}\right)\right] \nonumber \\ &=& - \int_{k(0)}^{k(t)}
  d\vec{k}' \cdot \vec{\mathcal{A}}_{k'},
  \label{Berry}
\end{eqnarray}
with the integral taken over the trajectory of $k(t)$.

\begin{figure}
\includegraphics[width=0.27\textwidth]{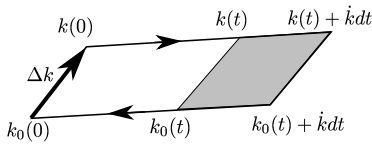}
\caption{$k$-space line integrals giving rise to the Berry's phases in
  equation (\ref{line integrals}).}
\label{k path}
\end{figure}

Now consider a wavepacket composed of Bloch states, initially centered on the
Bloch state with $k=k_0$:
\begin{eqnarray}
  \psi(r,0) &=& \sum_k f(|k-k_0|)\,u_k(r)\,e^{ikr} \\ &=& e^{ik_0r} \sum_k
  f(|\Delta k|)\,u_k(r)\,e^{i\Delta k \cdot r},
  \label{initial wavepacket}
\end{eqnarray}
where $\Delta k = k - k_0$ and $f$ is some envelope function.  From
(\ref{psit}) and (\ref{Berry}), the wavepacket at time $t$ is
\begin{multline}
  \psi(r,t) = e^{ik_0r} \sum_k f(|\Delta k|) \, u_{k(t)}(r) \, \\ \times
  e^{i\Delta k \cdot r} \exp\left[-\frac{i}{\hbar} \int_0^t dt'
    E(k(t'))\right]\, e^{i\gamma_{k}(t)}.
  \label{wavepacket}
\end{multline}
Note that $k(t) - k_0(t) = \Delta k$ is time-independent.  Comparing
(\ref{wavepacket}) to (\ref{initial wavepacket}), we observe
displacements in both $r$ and $k$.  The $k$-space displacement is
simply $q(t)$.  The $r$-space displacement is determined by the phase
factors on the last line of (\ref{wavepacket}), which should have the
form
\begin{equation*}
  (\textrm{overall phase factor}) \times \exp\left[i\Delta k \cdot
    r(t)\right],
\end{equation*}
where the overall phase factor comes from the phase of the central
wavepacket $k_o(t)$ and the second term comes from the phase
difference between $k(t)$ and $k_o(t)$.

From these considerations, we see that the phase differences between $k$ and
$k_0$ which arise from the band energies yield the usual group velocity:
\begin{eqnarray}
  \Delta k \cdot \vec{v}_g &=& \frac{d}{dt}\left\{\frac{1}{\hbar}
  \int_0^t dt' \left[E(k(t')) - E(k_0(t'))\right] \right\} \\ &=&
  \Delta k \cdot \left[\frac{1}{\hbar}\nabla_kE(k_0(t))\right].
  \label{line integrals}
\end{eqnarray}

I now claim that the Berry's phase differences give the anomalous velocity:
\begin{equation}
  \Delta k \cdot \vec{v}_a = \frac{d}{dt} \left\{ \int_{k(0)}^{k(t)}\!
  d\vec{k}'\cdot\vec{\mathcal{A}}_{k'} - \int_{k_0(0)}^{k_0(t)}\!
  d\vec{k}''\cdot\vec{\mathcal{A}}_{k''} \right\}.
\end{equation}
To prove this, observe that the line integrals are taken over the top and
bottom segments of the parallelogram in Fig.~\ref{k path}.  When $\Delta k$ is
sufficiently small, integrals $\int d\vec{k}'\cdot \vec{\mathcal{A}}_{k'}$
over the side segments become negligible; then the two separate line integrals
can be replaced with a single edge integral, taken clockwise around the
boundary $\Gamma(t)$ of the paralleogram:
\begin{eqnarray}
  \Delta k \cdot \vec{v}_a &\approx& \frac{d}{dt} \left\{ \oint_{\Gamma(t)}
  d\vec{k}'\cdot\vec{\mathcal{A}}_{k'} \right\} \\ &=& - \frac{d}{dt} \left\{
  \int_{a(t)} da \cdot \left(\nabla_k \times \vec{\mathcal{A}}_{k'} \right)
  \right\}.
\end{eqnarray}
In time $dt$, the edge advances by $\dot{k}dt$, and the additional area (grey
region in Fig.~\ref{k path}) is $(\dot{k} \times \Delta k) dt$.  Thus,
\begin{eqnarray}
  \Delta k \cdot \vec{v}_a &=& \left( \nabla_k \times
  \vec{\mathcal{A}}_{k_0}\right) \cdot \left(\Delta k \times \dot{k}\right)
  \\ &=& \Delta k \cdot \left(\dot{k} \times \nabla_k \times
  \vec{\mathcal{A}}_{k_0}\right).
\end{eqnarray}
This is precisely the anomalous velocity given in (\ref{rdot}).

I am grateful to P.~A.~Lee for helpful discussions.

\end{document}